\documentstyle[prl,aps,multicol,tighten,epsf]{revtex}

\begin{document}
\draft

\title{Short Time Behavior in De Gennes' Reptation Model}
\author{Ute Ebert$^{1,3}$, Artur Baumg\"artner$^2$, 
and Lothar Sch\"afer$^3$}
\address{
$^1$Instituut--Lorentz, Universiteit Leiden, Postbus 9506, 
2300 RA Leiden, the Netherlands,
         }
\address{
$^2$Institut f\"ur Festk\"orperforschung and Forum Modellierung, 
Forschungszentrum J\"ulich, 52425 J\"ulich, Germany,
         }
\address{
$^3$Fachbereich Physik, Universit\"at Essen, 45117 Essen, 
Germany.
         }
\date{January 14, 1997}
\maketitle

\begin{abstract} 
To establish a standard for the distinction of reptation from 
other modes of polymer diffusion, we analytically and numerically 
study the displacement of the central bead of a chain diffusing 
through an ordered obstacle array for times $t<{\cal O}(N^2)$. 
Our theory and simulations agree quantitatively and show
that the second moment approaches the $t^{1/4}$ often viewed as
signature of reptation only after a very long transient and only 
for long chains ($N>100$). Our analytically solvable  
model furthermore predicts a very short transient for the fourth  
moment. This is verified by computer experiment. \linebreak
\end{abstract}

\begin{multicols}{2}

Since the reptation model was formulated 25 years ago by 
De Gennes \cite{dG}, it has become a widely used concept of 
polymer motion. In its original form it is concerned with 
the diffusion of a mobile polymer chain through a net of
impenetrable and immobile obstacles. It describes this 
motion as being confined to a tube created by the surrounding 
obstacles. Diffusion takes place by the motion of 
little wiggles of surplus length, the ``defects'', along the tube. 
Asymptotically this scenario yields simple power laws for various 
quantities \cite{dG}. For instance, the diffusion constant $D$, 
which governs the motion of the total chain at large time scales, 
should behave as $D \propto N^{-2}$,
where $N$ is the chain length, i.e., the polymerization index of 
the macromolecule. Another important prediction concerns the 
motion of the central bead ${\bf r}_{N/2}(t)$. For intermediate 
times it is predicted to diffuse a mean square distance 
\begin{eqnarray}
\label{1}
g_{1}(t) &=& 
\left\langle ({\bf r}_{N/2}(t) - {\bf r}_{N/2}(0))^{2}\right\rangle 
\nonumber\\
&\propto& \left\{ \begin{array}{ll}
t^{1/4} & \;\; \mbox{ for } T_0 \ll t \ll T_2 \; \; \; \\
t^{1/2} & \;\; \mbox{ for } T_2 \ll t \ll T_3 \; \; \;. 
\end{array} \right.
\end{eqnarray}
Here $T_0={\cal O}(N^0)$ is a microscopic time and 
$T_2={\cal O}(N^2)$ is the Rouse time, i.e., the characteristic 
scale over which an unconstrained chain of length $N$ equilibrates. 
The ``reptation time'' $T_3={\cal O}(N^3)$ marks the onset of the 
asymptotic diffusional regime, where the central bead
just follows the diffusion of the center of mass: $g_{1}(t) = D t$.
For comparison, the Rouse model for a free noninteracting chain 
yields $g_{1}(t) \propto t^{1/2}$ for $t \ll T_{2}$ and asymptotic 
diffusion of both central bead and center of mass with diffusion 
constant $D \propto N^{-1}$ for $t \gg T_{2}$.

From the beginning, the reptation model has also been applied to 
polymer motion through melts, dense solutions or gels, even though 
in such systems the surrounding medium more or less can relax. 
A lot of experimental or computer experimental work has aimed
at verifying the reptation predictions in such systems. 
(See \cite{Lo} for a recent review.) 
The outcome of these efforts today is somewhat ambiguous. 
The expected power laws never have been established beyond doubt. 
Compared to the free motion of a chain one typically finds some 
slowing down giving rise to effective power laws, with no obvious
unique and simple interpretation \cite{KS,KG}. $g_1(t)$, 
for instance, typically behaves as 
\begin{eqnarray*}
g_{1}(t) \propto t^{x} \;\; \mbox{ with } 0.25 < x \lesssim 0.4 
\;\; \mbox{ for } T_0 \lesssim t \lesssim T_2~.
\end{eqnarray*}
It never is clear whether such deviations from predictions like (1)  
are due to the relaxation of the surrounding medium or are intrinsic 
to the reptation scenario in the available range of chain lengths. 
Furthermore it must be stressed that contrary to statements  
often found in the literature, also in a frozen disordered 
environment the asymptotic reptation behavior has not been properly 
observed. In fact, recent work \cite{SW,EB,St} suggests that even 
strictly immobile obstacles can ruin the reptational power laws, 
due to their disordered spatial distribution. This finding is in 
contrast to the view prevailing in the older literature. 
(See \cite{EE}, for instance.) 

In view of this situation it is somewhat surprising to note 
that little effort has been made to clarify the implications of the 
original reptation model, considering polymer diffusion in an 
{\em ordered} array of {\em fixed} obstacles. Clearly only such work 
can provide the basis for a controlled analysis of effects of disorder 
or relaxation of the surrounding medium. We are aware of only two 
such studies. In early work \cite{EE} Evans and Edwards claimed to find 
the behaviour (1), but the statistics of their data is insufficient 
for a convincing analysis. In fact, our extended study of their model 
as presented below clearly demonstrates that for their chain
lengths the $t^{1/4}$-law is not yet attained. More recently Deutsch  
and Madden \cite{DM} reconsidered the model and found $D \propto N^{-2.5}$ 
for $N \lesssim 100$, in contrast to the expected $N^{-2}$-law.

In order to establish the predictions of the pure reptation
model, that serves as a starting point for the investigation of more
complicated environments, 
we here present results of an extended study of reptation through 
an ordered array of fixed obstacles in three dimensions. Concentrating 
on the internal motion in a time regime prior to free diffusion 
we present new extensive simulations, and we introduce an analytically 
solvable model that modifies the original reptation model of De Gennes 
\cite{dG} to take the discreteness of the chain into account. We find 
quantitative agreement among our theory and our data. Simple power laws, 
however, are found only after a surprisingly long initial transient and 
are reached only by quite long chains. Our theory explains the long 
transient quantitatively as a consequence of the finite segment size. 
Furthermore our theory suggests to determine another quantity, 
not measured previously, which should show the reptational power 
laws more clearly, without the long transient. Also this 
prediction is verified by computer experiment. In summary, 
we provide an analytically understood and numerically tested 
standard for reptation of discrete chains of finite length, 
relevant for the interpretation of all previous simulation results,
and merging asymptotically ($N \to \infty$, $t\gg 1$) with the 
earlier predicted power laws (1) \cite{dG}. In future work, this should 
allow for a meaningful analysis of the influence of disorder or of 
some slow relaxation of the medium on reptation.

\begin{figure}[h]
\setlength{\unitlength}{1cm}
\begin{picture}(8,4.5)
\epsfxsize=10cm
\put(-.3,-9){\epsffile{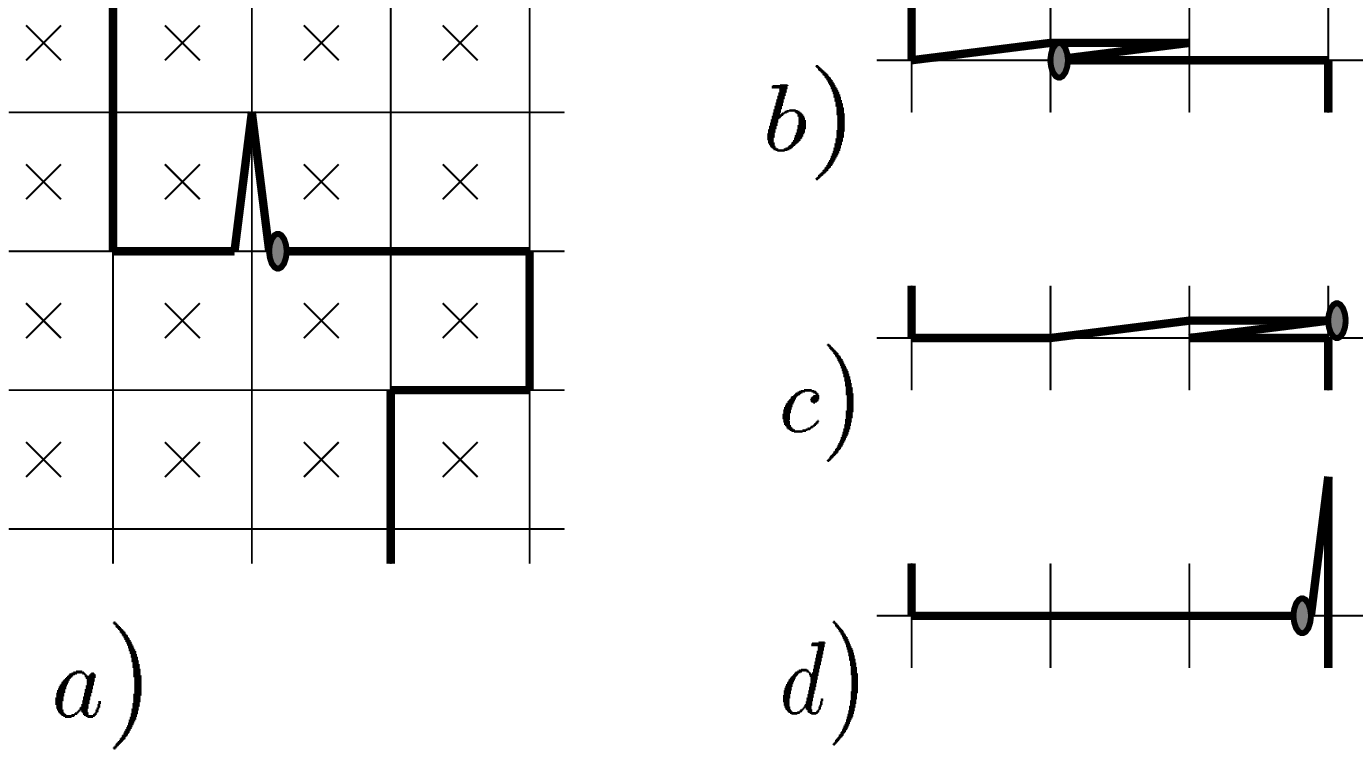}}
\end{picture}
\begin{center}
\begin{minipage}{8cm}
\small FIG.\ 1. The Evans-Edwards model:
$a)$ Section of a random walk on a square lattice, 
showing one hairpin. The crosses represent the obstacle lattice. 
The oval identifies a bead of the chain. While the hairpin 
diffuses past it, the bead is transported two steps along 
the tube towards configuration $d)$. $b)$ and $c)$ give 
intermediate configurations. For clearer representation 
we have opened the base of the hairpins.
\end{minipage}
\end{center}
\end{figure}

In our simulations we use the Evans-Edwards model \cite{EE}: 
The chain is confined to a cubic lattice. 
A second identical lattice is taken to represent the obstacles and 
is placed such that the lattice points coincide with the centers 
of the cells of the first lattice. Then the only moves allowed are 
those of ``hairpins'', which can rotate among lattice bonds (see 
Fig.\ 1). This model incorporates the smallest possible tube 
diameter. It therefore shows the strongest 
restriction of chain motion and should exhibit reptation in 
clearest form. In our numerical experiment we concentrated on 
the short-time behaviour (1), searching for the $t^{1/4}$-law. 
We note, however, that for shorter chains $(N \leq 100)$ we also 
covered times $t > T_{3}$ and recovered the results of \cite{DM}.
A full account of our results will be published elsewhere.

Fig.\ 2 shows $g_1(t)$ for times $t \lesssim T_2$. Obviously, the 
expected power law $g_{1}(t) \propto t^{1/4}$ indeed is attained, 
but only very slowly. It fully is developed only for surprisingly 
long chains $(N \gtrsim 100)$. It seems that for $N\to\infty$, there 
is a well defined limiting curve, that merges with the expected 
power law only for Monte Carlo time $t_{MC} \gtrsim 3 \cdot 10^{4}$.
($t_{MC}=1$ stands for one attempted move per bead of the chain.) 
For our longest chain $(N = 600)$ the $t^{1/4}$-power law regime 
then extends over about two decades. For $N \lesssim 100$ the 
influence of the hairpins diffusing in from the ends of the chain 
becomes important, i.e., $T_2$ is exceeded, before the power law 
properly is developed. The curves then bend upwards according to
Eq.\ (\ref{1}). Effective exponents $\frac{1}{4} < x \lesssim 0.4$ 
could easily be extracted for shorter chains, in full agreement with 
earlier observations. 

\begin{figure}[h]
\setlength{\unitlength}{1cm}
\begin{picture}(8,5.5)
\epsfxsize=20cm
\put(-3.7,-16){\epsffile{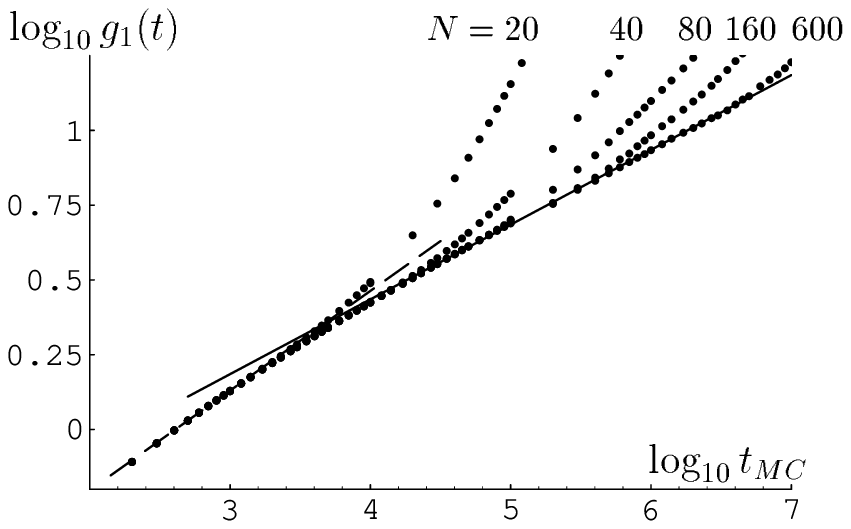}}
\end{picture}
\begin{center}
\begin{minipage}{8cm}
\small FIG.\ 2.
Doubly logarithmic plot of Monte Carlo data for various
chain lengths $N$. 
The straight lines give slope $1/4$ or $1/3$, respectively.
\end{minipage}
\end{center}
\end{figure}

To properly understand these results we worked out the quantitative 
predictions of the original reptation model \cite{dG} in a fully 
discretized version. The defects are modelled as noninteracting 
particles, freely diffusing along a lattice chain. At the chain ends 
defects can be destroyed or created, which is modelled by coupling 
the ends to large reservoirs. If a defect passes a bead, that bead 
is displaced by two steps along the ``tube'', that is defined by the 
$t=0$ configuration of the chain (see Fig.\ 1). Let 
$n_{\pm} (t,j)$ be the number of defects passing bead $j$ within 
time interval $t$ from the left ($+$) or right ($-$), respectively. 
Then $2 \cdot n (t,j)$ with
\begin{equation}
\label{2}
n (t,j) = n_{+}(t,j) - n_{-}(t,j)
\end{equation}
gives the displacement of that bead along the ``tube''. If $t$ is 
short compared to the tube renewal or reptation time 
$T_{3} = {\cal O}(N^{3})$, this effective diffusion for a central 
bead takes place in the fixed initial tube. Since this tube is 
a realization of a random walk on a cubic lattice with lattice 
constant $a = 1$, i.e., with lattice vectors ${\bf s}_i$ with 
cartesian components $s_i^\mu \in \{0, \pm 1\}$ and 
$\langle {\bf s}_i \cdot {\bf s}_{i'}\rangle_t = \delta_{i i'}$, 
the average over all tube configurations yields for $t \ll T_3$
(a lower index $c$ denotes the average over tube configurations
and $d$ over thermal defect dynamics)
\begin{eqnarray}
\label{3}
g_{1}(t) &=& \left\langle 
\Big({\bf r}_{N/2}(t) - {\bf r}_{N/2}(0)\Big)^2 
\right\rangle_{c,d}
\nonumber\\ 
&=& \left\langle 
  \left( \sum_{i=1}^{2|n|} {\bf s}_i \right)^2
  \right\rangle_{c,d}
 =  \left\langle \left|\; 2 \cdot n \left(t,\frac{N}{2}\right) 
    \right| \right\rangle_d~,         
\end{eqnarray}
a result that should hold up to a microstructure dependent additive
constant $c_0$ much less than one. Such corrections for instance 
arise from configurations where the bead at times $0$ or $t$
sits on the tip of a hairpin, such configurations being not 
included faithfully in our defect model. Note that we use the notion 
of the ``tube'' in a somewhat loose way, not distinguishing it 
properly from the chain configuration. In the MC-model the tube 
can be defined as the chain configuration with all hairpins cut off, 
resulting in a non-reversal random walk. The simple random-walk-like
spatial embedding {\bf r} of the internal chain coordinates implied 
by Eq.\ (\ref{3}) is justified by the observation that both defects 
and bead move along the chain, not along the tube. 

Our model is similar in spirit to the repton model proposed by  
Rubinstein \cite{Ru}, but somewhat closer to the original
De Gennes model. It allows to analytically 
calculate dynamical properties like the time dependent 
distribution function ${\cal P}(t;n,j)$ of $n(t,j)$, while 
taking the proper defect dynamics at the chain ends into account. 
Leaving details for a forthcoming paper, we only present 
some results for the central bead $j = N/2$. The second 
moment of the distribution of $n(t,N/2)$ is found as
\begin{eqnarray}
\label{4}
\lefteqn{
\left\langle  \;n^2 \left(t,\frac{N}{2}\right)  \right\rangle_d = 
\rho_0\;\Bigg[
\frac{2\:t}{N} + \frac{N}{6} + \frac{1}{3N}
         }
\\
& & \qquad\quad + \;\frac{1}{N} \sum_{k=1}^{N-1} (-1)^{k} 
\left(1 - \frac{\cos^2\frac{\pi k}{2}}{\sin^2\frac{\pi k}{2 N}}
\right) \;
\mbox{\large e}^{\textstyle - 4 t\: \sin^2\frac{\pi k}{N}}
\Bigg]~,
\nonumber 
\end{eqnarray}
where $\rho_0$ is the density of defects. In Eq.\ (\ref{4}) we 
recognize the sum over the Rouse modes of the chain, typical for 
such problems. $g_{1}(t)$ [Eq.\ (\ref{3})] is found as a function 
of $\langle n^{2} \rangle$ as
\begin{eqnarray}
\label{5}
g_{1}(t) &=& \langle |2n|\rangle =
\frac{\left( 2 \langle n^{2} \rangle\right)^{1/2}}{\pi} \Bigg[ 
2\sqrt{\pi} + \Gamma \left(- \frac{1}{2},
2 \langle n^{2} \rangle\right) 
\\
& & - \int_0^{2 \langle n^2\rangle} 
\!\!\! dx\:x^{-3/2} e^{-x}
\left( \left(1 - \frac{x}{2 \langle n^2\rangle}\right)^{-1/2} 
\!\!\!- 1 \right) \Bigg] ~,
\nonumber 
\end{eqnarray}
where $\Gamma(x,y)$ denotes the incomplete Gamma function. 
In Eq.\ (\ref{5}) both the term $\Gamma(\ldots)$ and the integral 
are due to the discrete structure of the chain. Evaluating these 
results in the time regime $T_0 \ll t \ll T_{2}$, that
mathematically is defined by the limit $N \to \infty$ and
$1 \ll t < \infty$ fixed, we find the intermediate asymptotics 
\begin{equation}
\label{6}
g_{1,\infty}(t)=4\;\pi^{-3/4}\;\rho_0^{1/2}\;t^{1/4}\; ~.
\end{equation}
In the time regime $T_2 \ll t \ll T_3$, defined as limit 
$N \to \infty$, $1 \ll t/N^2 < \infty$ fixed, we recover 
the $t^{1/2}$-behavior: $g_1(t) = 4 \;(\rho_0/\pi)^{1/2}\; 
(t/N)^{1/2}$. These are the well known asymptotic reptation 
predictions (1).

In comparing our quantitative results to the simulation data,
we have to specify the density of defects $\rho_0$. Due to the 
occurence of larger side loops (double hairpins etc.), the mapping
of hairpins onto defects is not one-to-one, so that $\rho_0$
is some effective model parameter. From the equilibrium statistics
of a random walk it is found to be fixed within very close bounds: 
$\frac{1}{9} \lesssim \rho_0 \lesssim \frac{1}{4}$. The shape of our 
theoretical curves is not sensitive to the precise value chosen, 
which mainly can be absorbed into the time scale. We here take 
$\rho_0 = 1/4$. We also need to fix the additive constant $c_0$ 
introduced above. From analyzing the data for microscopic times 
$(t_{MC} \lesssim 30)$, we choose to subtract $c_0 = 0.1$ from the
Monte Carlo data for $g_1(t)$. The only really free parameter is the 
scale $\tau = t/t_{MC}$, relating the time variable $t$ of the 
analytical model to the Monte Carlo time $t_{MC}$. By adjusting 
this single para\-meter to $\tau = 0.01$, we find the results of 
Fig.\ 3. Note that since we devided out the intermediate 
asymptotics (6), this plot is much more sensitive to small deviations 
than the usual doubly logarithmic representation. In view of this 
we believe the agreement found for $N \geq 40$ to be truly remarkable.

\begin{figure}[h]
\setlength{\unitlength}{1cm}
\begin{picture}(8,6)
\epsfxsize=20cm
\put(-3.5,-16){\epsffile{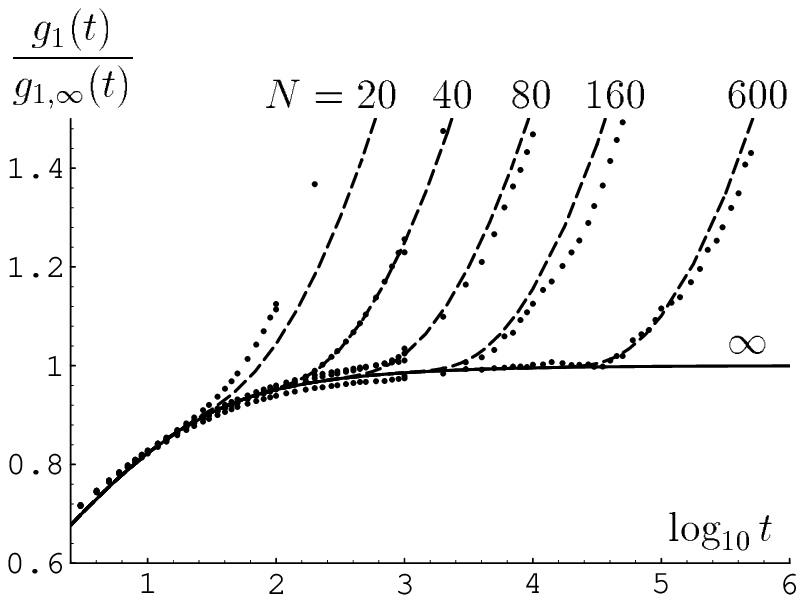}}
\end{picture}
\begin{center}
\begin{minipage}{8cm}
\small FIG.\ 3.
$g_1(t)/g_{1,\infty}(t)$ plotted against $\log_{10}t$,
$t=0.01 \cdot t_{MC}$. Full and broken lines: theory.
Points: MC data.
\end{minipage}
\end{center}
\end{figure}

We now can analyze in detail the origin of the long transient, 
that in Fig.\ 3 extends up to $t \approx 10^{3}$. According 
to Eq.\ (\ref{5}), $g_1(t)$ depends on $t$ only via 
$\langle n^{2}(t,N/2)\rangle$. Now it is found that
$\langle n^{2}(t,N/2)\rangle$ approaches its intermediate 
asymptotics $\langle n^{2}(t,N/2)\rangle^{1/2} = 
(2 \rho_0)^{1/2}(t/\pi)^{1/4}$ for times as small as 
$t \gtrsim 3$. The slow transient thus is due to the 
deviations of $\langle |n|\rangle$ from $\big(2\langle 
n^{2}\rangle/\pi\big)^{1/2}$ for small $\langle n^2\rangle$
as exhibited in Eq.\ (\ref{5}). These deviations reflect
the discrete structure of the chain. A simple toy model
explains this effect: The fluctuating quantity $n(t,j)$
takes only integer values and its distribution rapidly
looks like a {\em discretized} Gaussian. The width of this
Gaussian, however, increases but slowly: 
$\langle n^{2}(t,j)\rangle^{1/2} \sim t^{1/4}$. Even for 
$t \approx 10^{3}$ it shows nonzero weight essentially only 
for $| n | \lesssim 5$. Thus the discrete nature of the 
fluctuating variable $n(t,j)$ stays important for a very large 
initial time interval. Evaluating $\langle | n(t,j) | \rangle$ 
with such a discretized Gaussian we recover the slow transient.

These considerations immediately suggest to eliminate the 
transient behaviour by measuring $\langle n^{2} \rangle$ 
instead of $\langle | n | \rangle$. Now for the present 
model it is easily checked that the cubic invariant 
$\hat{g}_1^2(t)$ reduces to $\langle n^{2} \rangle$: 
\begin{eqnarray}
\label{7}
\hat{g}_1(t) 
&=&  \left\langle \;\sum^{3}_{\mu = 1}\:
\left(r_{N/2}^\mu(t) - r_{N/2}^\mu(0)\right)^4 
\right\rangle_{c,d}^{1/2}\\
&=&  \left\langle \;\sum_{\mu=1}^3
     \left( \sum_{i=1}^{2|n|} s_i^\mu \right)^4
     \right\rangle_{c,d}^{1/2}
= \left\langle \;4 \cdot n^2\left(t,\frac{N}{2}\right) 
  \right\rangle_d^{1/2}~.
\nonumber
\end{eqnarray}
For random-walk-type chains on other lattices or in 
the continuum the spatial embedding {\bf r} of the basic dynamic
quantity $\langle n^2 \rangle$ differs, but $\langle n^2 \rangle$
always can be expressed by an appropriate combination of the fourth 
and the second moment of $({\bf r}_{N/2}(t) - {\bf r}_{N/2}(0))$. 
Fig.\ 4 shows a doubly logarithmic plot of our data for $\hat{g}_{1}$ 
as compared to $g_{1}$ for chain lengths $N$ = 40 and 600. 
Clearly our expectation is born out. For $\hat{g}_{1}(t)$ only
some slight initial deviation from the $t^{1/4}$ power law can 
be detected, that could lead to a tiny underestimation of the 
exponent $( \approx 0.235$ in place of $0.25$). 
Even for $N = 40$ we find a decent $t^{1/4}$-regime, that for 
$N = 600$ extends over almost four decades. We thus conclude
that our fourth moment shows asymptotic reptational 
behaviour much more clearly than the commonly used second moment.

\begin{figure}[h]
\setlength{\unitlength}{1cm}
\begin{picture}(8,5)
\epsfxsize=20cm
\put(-3.5,-16){\epsffile{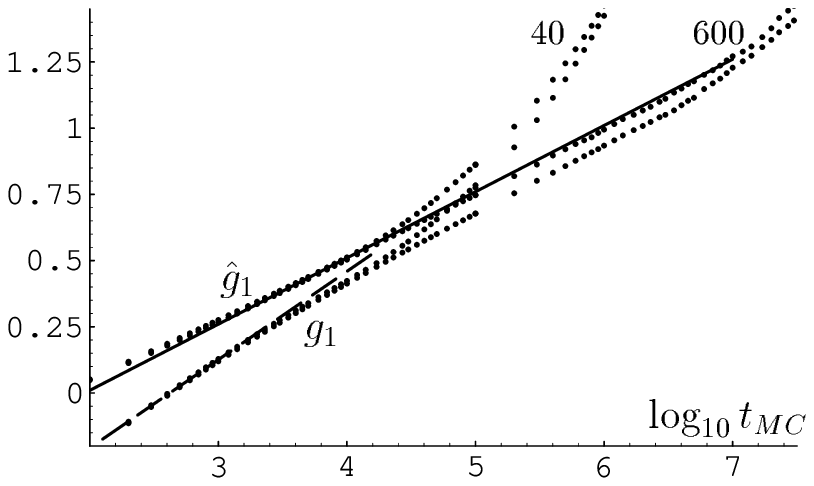}}
\end{picture}
\begin{center}
\begin{minipage}{8cm}
\small FIG.\ 4.
Doubly logarithmic plot of MC-data (like in Fig.\ 2)
for $g_1(t)$ and $\hat{g}_1(t)$ and for $N$ = 40, 600. 
Full line: $t^{1/4}$-law. Broken line: $t^{1/3}$.
\end{minipage}
\end{center}
\end{figure}

In summary, combining a simple exactly solvable model with computer
experiments, we not only recovered the asymptotic reptation
prediction of a $t^{1/4}$ regime, but we also quantitatively 
explained the shape of the crossover function with its long initial 
transient. Since any change of the microstructure of the model like 
an enlargement of the tube radius will only increase the initial 
effects, we believe our results to be typical also for other models. 
We have demonstrated that an appropriate fourth moment measuring 
$\langle n^2 \rangle$ instead of $\langle |n| \rangle$, suppresses 
the initial transient and thus lends itself to a much simpler 
analysis. These results are hoped to provide a standard for the
study of deviations from reptation due to disorder or due to 
a relaxation of the tube. In all cases we strongly recommend to
include the fourth moment of the displacement of the central bead
into the analysis.   

We thank J.M.J.\ van Leeuwen and D.\ Aalberts for helpful 
discussions. UE was supported by the Dutch Science Foundation NWO.
Support by the SFB 237 of the German Science Foundation DFG is also 
gratefully acknowledged. The simulations have been performed at
the Forschungszentrum J\"ulich.

\end{multicols}

\end{document}